\documentclass[sigconf,natbib=true,anonymous=false]{acmart}
\AtBeginDocument{%
  }

\usepackage{multirow}

\begin{document}

\title{MIDSim: Simulating Multi-Channel Information Diffusion in Social Media with LLM-Powered Multi-Agent System}

\author{Lexi Liu}
\affiliation{%
  \institution{State Key Laboratory of AI Safety, Institute of Computing Technology, Chinese Academy of Sciences}
  \institution{University of Chinese Academy of Sciences, Beijing,China}
  \country{}}
\email{liulexi2004@outlook.com}

\author{Qi Cao}
\authornote{Corresponding author.}
\affiliation{%
  \institution{State Key Laboratory of AI Safety, Institute of Computing Technology, Chinese Academy of Sciences}
  \city{Beijing}
  \country{China}}
\email{caoqi@ict.ac.cn}

\author{Yuanhao Liu}
\affiliation{%
  \institution{State Key Laboratory of AI Safety, Institute of Computing Technology, Chinese Academy of Sciences}
  \city{Beijing}
  \country{China}}
\email{liuyuanhao20z@ict.ac.cn}

\author{Huawei Shen}
\authornotemark[1]
\affiliation{%
  \institution{State Key Laboratory of AI Safety, Institute of Computing Technology, Chinese Academy of Sciences}
  \city{Beijing}
  \country{China}}
\email{shenhuawei@ict.ac.cn}

\author{Xueqi Cheng}
\affiliation{%
  \institution{State Key Laboratory of AI Safety, Institute of Computing Technology, Chinese Academy of Sciences}
  \city{Beijing}
  \country{China}}
\email{cxq@ict.ac.cn}

\renewcommand{\shortauthors}{Liu et al.}

\begin{abstract}
Information diffusion in social media shapes public opinion and collective behavior, making its modeling and simulation an important research problem.
Existing studies have investigated information diffusion through epidemic-based, cascade-based, and point process models. 
However, they predominantly focus on diffusion through social links, overlooking other diffusion channels enabled by platform algorithms (e.g., recommender systems) and failing to capture user behavioral complexity.
To address these limitations, we propose an LLM-powered multi-agent system for simulating multi-channel information diffusion, where large language models instantiate personalized user agents and the diffusion process jointly models social and algorithmic exposure streams. 
We further construct three real-world diffusion dataset spanning Sina Weibo, RedNote, and Twitter, containing diffusion records, user profiles, historical posts, and social relationships. 
Experimental results on real diffusion events show that our proposed framework realistically simulate macro diffusion phenomenon and generate diverse comment content, significantly outperforming baselines.
\end{abstract}

\ccsdesc[500]{Human-centered computing~Social media}

\keywords{Information Diffusion, Multi-Agent Systems, Diffusion Simulation}

\maketitle

\section{Introduction}
Social media platforms have become major infrastructures for information dissemination, shaping the emergence, spread, and evolution of opinions. Understanding how information propagates across these platforms is essential for studying collective behaviors and social dynamics. Consequently, the modeling and simulation of information diffusion have attracted extensive research attention.

\begin{figure*}[h]
  \centering
  \includegraphics[width=0.88\linewidth]{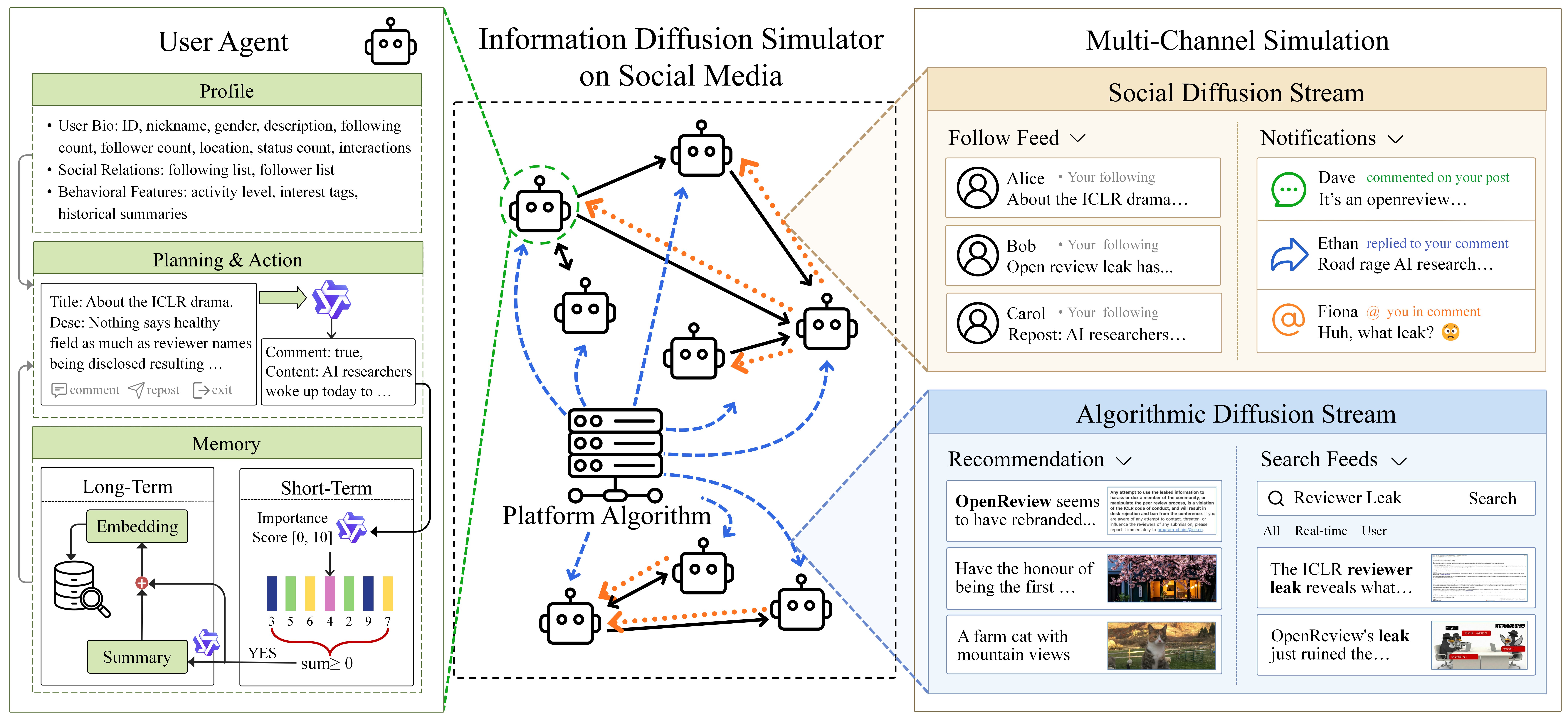}
  \caption{Framework of MIDSim, consisting of user agents and multi-channel diffusion simulation. Orange dashed lines represent social diffusion streams, whereas blue dashed lines represent algorithmic diffusion streams.}
  \label{fig:architecture}
\end{figure*}

Existing studies have investigated information diffusion through a variety of modeling paradigms, including epidemic-based models~\cite{liu2024review,govindankutty2024epidemic}, point process models~\cite{zhang2022anytime,jing2025your}, and cascade-based models~\cite{goldenberg2001talk,granovetter1978threshold,chen2011influence,kempe2003maximizing,zhang2024information,zhang2024neural,bhagat2012maximizing}. While epidemic-based and point-process models are effective for characterizing aggregate diffusion dynamics, they provide limited insight into the underlying propagation process. In contrast, cascade-based models, such as Independent Cascade (IC)~\cite{goldenberg2001talk,chen2011influence,zhang2024information,zhang2024neural} and Linear Threshold (LT)~\cite{granovetter1978threshold,kempe2003maximizing,bhagat2012maximizing}, explicitly model diffusion as a propagation process over social networks, providing an intuitive and interpretable framework.

Despite their success, cascade-based models assume that information diffusion is primarily governed by social links. However, on modern social media platforms, users increasingly discover and consume information through algorithmic channels, such as recommendation and research feeds, in addition to their social networks. This shift challenges the assumption underlying existing models, limiting their ability to faithfully characterize real-world diffusion processes. Moreover, users are often represented by simplified propagation probabilities, overlooking the behavioral complexity and diversity observed in real-world information diffusion.

To solve these limitations, we propose an LLM-powered multi-agent system for \underline{m}ulti-channel \underline{i}nformation \underline{d}iffusion \underline{sim}ulator (MIDSim). Leveraging the strong behavioral simulation capabilities of LLMs~\cite{gao2023s3,tornberg2023simulating,guozhen2024human,guo2024large,li2024econagent,mou2024unveiling}, MIDSim represents social media users as personalized LLM-powered agents grounded in their profiles and historical behaviors, rather than simplified behavioral assumptions. Furthermore, MIDSim models information exposure through both social channels (e.g., follow feeds and notifications) and algorithmic channels (e.g., recommendation and search feeds). Together, these designs provide a unified framework for realistically simulating information diffusion on modern social media platforms.

To enable realistic evaluation, we construct three real-world diffusion datasets spanning Sina Weibo, RedNote, and Twitter, containing diffusion records, user profiles, historical posts, and social relationships. 
Experimental results on these real diffusion events show that our proposed MIDSim not only accurately reproduces macro diffusion phenomenon but also generates realistic and diverse user comments, significantly outperforming baselines. These results highlight the importance of modeling information diffusion as a multi-channel process while capturing complex user behaviors through personalized agents.

\section{Methodology}
As shown in Figure~\ref{fig:architecture}, MIDSim simulates the diffusion of an observed social media event with LLM-based user agents and multi-channel diffusion simulation. 
Starting from event-related posts, MIDSim simulates the evolution of an event time by step. At each step, agents are exposed to content through social and algorithmic channels, decide whether and how to engage with it, and generate new content that becomes part of the evolving diffusion process.

\subsection{User Agent}
Different from simplified probabilistic diffusion models such as IC and LT, MIDSim represents social media users as personalized agents following multi-agent system frameworks~\cite{park2023generative,yan2024opencity,piao2025agentsociety,lin2025simspark}. Given an agent set $A=\{a_1,a_2,\ldots,a_N\}$, each agent $a_i$ corresponds to one user and makes diffusion-related decisions based on its profile, memory, current observation, and available actions.

\subsubsection{Profile.}
For real-data-driven simulation, each profile is initialized from observable platform data, including user bio and social relations, as well as behavioral features such as activity level, interest tags, and historical summaries. 
The activity level is computed as $act_i=\operatorname{Norm}(\log(1+x_i))$, where $x_i$ denotes the number of interactions or statuses of user $i$. 
For lengthy posts and followed accounts, we use Qwen~\cite{hui2024qwen2} to generate compact interest tags and historical summaries for agent personalization.

\subsubsection{Memory and Decision-making.}
Following common designs, each agent maintains recent interaction records and summarized long-term memories. 
Relevant memories are retrieved with BGE embeddings~\cite{xiao2024c} and combined with the profile and current observation as the decision context. 
The agent then forms a behavioral intention and selects actions from the action space.

\subsubsection{Action Space.}
The action space includes browsing content from exposure streams, deciding whether and how to propagate content, generating personalized text, interacting with eligible users, and maintaining attention to subsequent related content.

\subsection{Algorithm Module}
The algorithm module models platform-algorithm-driven exposure, which complements social diffusion by delivering content beyond users' direct social links. It contains two simplified components: recommendation for feed-based passive exposure and search for user-initiated content retrieval.

\subsubsection{Recommendation Algorithm.} 
We do not aim to reproduce a specific platform recommender. Instead, we approximate recommendation exposure with an LLM-based interest matching and reranking procedure. Since the complete platform content space is inaccessible, we construct a substitute global content pool from two sources: posts sampled from platform-specific interest categories to cover users' general interests, and fine-grained candidates from topics adjacent to the target topic to capture event-related interests. Before simulation, each user is assigned an initial candidate list from this pool. To handle the LLM context limit, we apply a hierarchical filtering to reduce the candidate space. During simulation, the reduced candidate pool is combined with target-topic-related content and reranked by the LLM according to interest relevance and content quality, forming the final recommendation list.

\subsubsection{Search Algorithm.}
Search exposure is implemented as a standard dense retrieval process. Given a query, we encode the query and candidate posts into dense representations, rank candidates by cosine similarity, and return the top-ranked posts as search results.

\subsection{Multi-Channel Information Diffusion}
MIDSim provides a simulation environment for multi-channel information diffusion. It manages the injection, exposure, propagation, and state updates of event-related information over time, thereby capturing the temporal evolution of diffusion under the joint effects of social relations and platform mechanisms. At each simulation step, agents are sampled as online or offline according to their activity levels; only online agents can receive exposed content and participate in propagation.

\subsubsection{Social Diffusion Stream.}
The social stream models exposure from observed social links and direct interactions. It includes follow-feed exposure, where online agents sample recent and previously unseen posts from followed users with probability $p_s$, and notification exposure, where replies, comments, and mentions are delivered to the target agents. When notifications exceed the attention budget $B_{\text{att}}$, the simulator samples a subset for processing.

\subsubsection{Algorithmic Diffusion Stream.}
The algorithmic stream models platform-mediated exposure beyond observed social links, including recommendation feed and search results. Recommendation feed represents passive system-pushed exposure: online agents receive recommended content with probability $p_a$, and the recommendation module selects previously unviewed posts from the candidate set. To reflect variable feed length and bounded attention, the number of recommended items $L$ follows a truncated geometric distribution with maximum length $K$:
\begin{equation}
P(L=l)=
\begin{cases}
(1-\alpha)\alpha^{l-1}, & 1 \le l < K,\\
\alpha^{K-1}, & l=K.
\end{cases}
\label{eq:rec_length}
\end{equation}
Search represents active information-seeking exposure. When the current observation is insufficient for decision-making, the agent is triggered to generate a query, and the algorithm retrieves relevant posts as supplementary exposure.

\subsubsection{Simulation Scheduling.}
The diffusion process advances in discrete but non-uniform time steps. Real diffusion events are typically bursty, with dense posts and feedback in the early stage and sparse activity in the later long-tail stage. Uniform time division would merge many early events into a few steps and delay feedback that could influence subsequent diffusion. MIDSim therefore uses finer temporal granularity in the early stage and longer windows in the later stage. For step $t$, its weight, duration, and cumulative boundary are defined as
\begin{equation}
w_t=e^{\rho t},\quad
\Delta\tau_t=\frac{D w_t}{\sum_{i=1}^{T}w_i},\quad
\tau_t=\sum_{k=1}^{t}\Delta\tau_k,
\label{eq}
\end{equation}
where $t=1,\dots,T$, $\rho>0$, $D$ is the total diffusion period, and $\tau_0=0$. Since $w_t$ increases with $t$, early steps correspond to shorter real-time windows and later steps to longer ones. 
The online probability is linearly mapped from the activity level. Thus, more active users are more likely to participate in exposure and propagation.

\section{Experiments}
This section describes the datasets, experimental settings, simulation quality, ablation studies and and case study.

\subsection{Data Collection}
To enable realistic implementation and evaluation, we construct three real-world diffusion datasets from Sina Weibo, RedNote, and Twitter, as existing public datasets often lack detailed information required for personalized agent construction and diffusion simulation. Data collection is conducted in accordance with relevant privacy regulations and platform policies. 

Centered on the ICLR 2026 reviewer leak event that began on November 27, 2025, we collect event-related posts through keyword search and record platform-specific interactions, including reposts on Sina Weibo, comments on RedNote, and quotes, retweets, replies on Twitter. We further parse @mentions from interaction content to enrich diffusion records. For users involved in the diffusion process, we collect profiles, historical posts, and follow relations. To mitigate the impact of incomplete social links, we exclude RedNote users whose social relations are not publicly accessible. Dataset statistics are reported in Table~\ref{tab:dataset_statistics}.

\subsection{Experimental Settings}

\begin{table}[t]
   \caption{Statistics of dataset.}
  \vspace{-0.5em}
  \label{tab:dataset_statistics}
  \begin{tabular}{lrrrr}
    \toprule
    Platform  & \#Propagations & \#Posts & \#Users & \#Relations \\
    \midrule
    Sina Weibo  & 114 & 23 & 130 & 209 \\
    RedNote  & 1,257 & 110 & 1,476 & 1,933 \\
    Twitter  & 1,674 & 44 & 1,067 & 25,643 \\
    \bottomrule
  \end{tabular}
\end{table}

\subsubsection{Baselines.}
We compare MIDSim with $7$ representative baselines, including IC-based models (IC~\cite{goldenberg2001talk}, IC-M~\cite{chen2012time}, IC-N~\cite{chen2011influence}, ICI~\cite{zhang2024information}) and LT-based models (LT~\cite{granovetter1978threshold}, LT-C~\cite{bhagat2012maximizing}, F-TM~\cite{barbieri2014influence}). 
Following common practice~\cite{zhang2024information}, the propagation probability or influence weight of edge $(u,v)$ is set to $1/|N_v^{in}|$, with other parameters kept as default. The result is obtained from $1000$ Monte Carlo simulations, and both the mean and standard deviation are reported.

\begin{table*}[t]
\centering
\caption{Prediction performance of different models (best results in bold and second-best results in italics).}
\vspace{-0.5em}
\label{tab:diffusion_baselines}
\resizebox{0.7\textwidth}{!}{
\begin{tabular}{llccccccccc}
\toprule
\multirow{2}{*}{Platform} & \multirow{2}{*}{Metric} 
& \multicolumn{4}{c}{IC-based models} 
& \multicolumn{3}{c}{LT-based models} 
& \multicolumn{2}{c}{MIDSim} \\
\cmidrule(lr){3-6} \cmidrule(lr){7-9} \cmidrule(lr){10-11}
& & IC & IC-M & IC-N & ICI & LT & LT-C & F-TM & Qwen2.5-14B & Llama-3.1-8B \\
\midrule

\multirow{2}{*}{Sina Weibo}
& MAPE 
& $0.586_{\scriptscriptstyle \pm 0.044}$ 
& $0.593_{\scriptscriptstyle \pm 0.044}$ 
& $0.670_{\scriptscriptstyle \pm 0.058}$ 
& $0.662_{\scriptscriptstyle \pm 0.036}$ 
& $0.499_{\scriptscriptstyle \pm 0.052}$ 
& $0.625_{\scriptscriptstyle \pm 0.046}$ 
& $0.600_{\scriptscriptstyle \pm 0.041}$ 
& $\mathbf{0.047_{\scriptscriptstyle \pm 0.041}}$ 
& $\underline{0.070_{\scriptscriptstyle \pm 0.040}}$ \\

& MRSE 
& $0.345_{\scriptscriptstyle \pm 0.051}$ 
& $0.354_{\scriptscriptstyle \pm 0.053}$ 
& $0.453_{\scriptscriptstyle \pm 0.079}$ 
& $0.441_{\scriptscriptstyle \pm 0.047}$ 
& $0.251_{\scriptscriptstyle \pm 0.052}$ 
& $0.392_{\scriptscriptstyle \pm 0.058}$ 
& $0.362_{\scriptscriptstyle \pm 0.049}$ 
& $\mathbf{0.003_{\scriptscriptstyle \pm 0.003}}$ 
& $\underline{0.006_{\scriptscriptstyle \pm 0.005}}$\\

\midrule

\multirow{2}{*}{RedNote}
& MAPE 
& $0.821_{\scriptscriptstyle \pm 0.009}$ 
& $0.835_{\scriptscriptstyle \pm 0.009}$ 
& $0.858_{\scriptscriptstyle \pm 0.013}$ 
& $0.853_{\scriptscriptstyle \pm 0.008}$ 
& $0.783_{\scriptscriptstyle \pm 0.008}$ 
& $0.837_{\scriptscriptstyle \pm 0.008}$ 
& $0.825_{\scriptscriptstyle \pm 0.008}$ 
& $\underline{0.037_{\scriptscriptstyle \pm 0.021}}$ 
& $\mathbf{0.022_{\scriptscriptstyle \pm 0.017}}$  \\

& MRSE 
& $0.675_{\scriptscriptstyle \pm 0.015}$ 
& $0.697_{\scriptscriptstyle \pm 0.015}$ 
& $0.736_{\scriptscriptstyle \pm 0.023}$ 
& $0.728_{\scriptscriptstyle \pm 0.014}$ 
& $0.613_{\scriptscriptstyle \pm 0.013}$ 
& $0.701_{\scriptscriptstyle \pm 0.014}$ 
& $0.681_{\scriptscriptstyle \pm 0.015}$ 
& $\underline{0.002_{\scriptscriptstyle \pm 0.002}}$ 
& $\mathbf{0.001_{\scriptscriptstyle \pm 0.001}}$ \\

\midrule

\multirow{2}{*}{Twitter}
& MAPE 
& $0.619_{\scriptscriptstyle \pm 0.097}$ 
& $0.928_{\scriptscriptstyle \pm 0.013}$ 
& $0.738_{\scriptscriptstyle \pm 0.080}$ 
& $0.814_{\scriptscriptstyle \pm 0.045}$ 
& $0.294_{\scriptscriptstyle \pm 0.224}$ 
& $0.422_{\scriptscriptstyle \pm 0.203}$ 
& $0.218_{\scriptscriptstyle \pm 0.033}$ 
& $\mathbf{0.024_{\scriptscriptstyle \pm 0.028}}$ 
& $\underline{0.031_{\scriptscriptstyle \pm 0.027}}$ \\

& MRSE 
& $0.393_{\scriptscriptstyle \pm 0.120}$ 
& $0.861_{\scriptscriptstyle \pm 0.024}$ 
& $0.552_{\scriptscriptstyle \pm 0.116}$ 
& $0.665_{\scriptscriptstyle \pm 0.071}$ 
& $0.137_{\scriptscriptstyle \pm 0.163}$ 
& $0.220_{\scriptscriptstyle \pm 0.187}$ 
& $0.049_{\scriptscriptstyle \pm 0.015}$ 
& $\mathbf{0.001_{\scriptscriptstyle \pm 0.002}}$ 
& $\underline{0.001_{\scriptscriptstyle \pm 0.002}}$\\

\bottomrule
\end{tabular}
}
\end{table*}

\subsubsection{Evaluation Metrics.}
We evaluate MIDSim on both diffusion prediction and text generation. For diffusion prediction, we adopt the commonly used MAPE~\cite{jing2025casft,cheng2024information} and MRSE~\cite{cao2020popularity,zhang2024information}. For text generation, following established practice~\cite{guo2024curious}, we evaluate the diversity from lexical (TTR~\cite{johnson1944studies}, Distinct-n~\cite{li2016diversity} with $n=3$, and Self-BLEU~\cite{zhu2018texygen}), semantic (Div$_{sem}$~\cite{guo2024curious}), syntactic (Div$_{syn}$~\cite{guo2024curious}) perspectives, and evaluate fidelity from text embedding similarity~\cite{xiao2024c}. Metrics are computed at the post level and then averaged.

\subsubsection{Implementation.}
MIDSim is built on YuLan-OneSim~\cite{wang2025yulan}, using Qwen2.5-14B-Instruct~\cite{hui2024qwen2} and Llama-3.1-8B-Instruct~\cite{grattafiori2024llama} as backbone LLMs (temperature $= 0$). BGE-1.5~\cite{xiao2024c} is adopted for embeddings. We set $B_{\text{att}}=10$, $K=5$, $\alpha \in [0.2,0.5]$, with $\rho=1.6$, $T=8$ and $D=24$ for scheduling; $(p_s,p_a)=(1,1)$ for Sina Weibo/RedNote and $(0.055,1)$ for Twitter. Each setting is evaluated over $3$ independent runs, with mean and standard deviation reported. Implementation details are available via the anonymous link\footnote{https://anonymous.4open.science/r/MIDSim-anonymous/}.

\subsection{Simulation Quality}
\subsubsection{Prediction Performance}
Table~\ref{tab:diffusion_baselines} reports the prediction performance of total diffusion volumes. MIDSim achieves the lowest errors across all platforms under all evaluation metrics, significantly outperforming representative IC- and LT-based baselines.
We also observe that, while MIDSim delivers consistently strong performance, simulation quality remains influenced by the choice of backbone LLM. Generally, Qwen2.5-14B achieves better results than Llama-3.1-8B, possibly due to differences in their underlying reasoning and instruction-following capabilities.

\begin{table}[t]
\centering
\caption{Diversity and fidelity of generated text.}
\label{tab:linguistic_diversity}
\vspace{-0.5em}
\resizebox{\columnwidth}{!}{%
\begin{tabular}{llcccccc}
\toprule
Platform & Setting & TTR & Distinct & Self-BLEU & Div$_{sem}$ & Div$_{syn}$ & Similarity\\
\midrule

\multirow{3}{*}{Sina Weibo}
& Real Data   & 0.841 & 0.998 & 0.991 & 0.562 & 0.738 & 1.000\\
\cmidrule(lr){2-8}
& w/o Profile & $0.351_{\scriptscriptstyle \pm 0.052}$ & $0.370_{\scriptscriptstyle \pm 0.064}$ & $0.109_{\scriptscriptstyle \pm 0.154}$ & $0.026_{\scriptscriptstyle \pm 0.037}$ & $0.048_{\scriptscriptstyle \pm 0.069}$ & $0.425_{\scriptscriptstyle \pm 0.004}$\\
& w/ Profile  & $\mathbf{0.561_{\scriptscriptstyle \pm 0.022}}$ & $\mathbf{0.748_{\scriptscriptstyle \pm 0.045}}$ & $\mathbf{0.852_{\scriptscriptstyle \pm 0.052}}$ & $\mathbf{0.336_{\scriptscriptstyle \pm 0.039}}$ & $\mathbf{0.382_{\scriptscriptstyle \pm 0.049}}$ & $\mathbf{0.454_{\scriptscriptstyle \pm 0.011}}$\\
\midrule

\multirow{3}{*}{RedNote}
& Real Data   & 0.722 & 0.964 & 0.983 & 0.540 & 0.708 & 1.000\\
\cmidrule(lr){2-8}
& w/o Profile & $0.147_{\scriptscriptstyle \pm 0.007}$ & $0.163_{\scriptscriptstyle \pm 0.007}$ & $0.262_{\scriptscriptstyle \pm 0.009}$ & $0.016_{\scriptscriptstyle \pm 0.003}$ & $0.029_{\scriptscriptstyle \pm 0.004}$ & $0.427_{\scriptscriptstyle \pm 0.001}$\\
& w/ Profile  & $\mathbf{0.416_{\scriptscriptstyle \pm 0.004}}$ & $\mathbf{0.730_{\scriptscriptstyle \pm 0.007}}$ & $\mathbf{0.298_{\scriptscriptstyle \pm 0.003}}$ & $\mathbf{0.431_{\scriptscriptstyle \pm 0.006}}$ & $\mathbf{0.347_{\scriptscriptstyle \pm 0.006}}$ & $\mathbf{0.467_{\scriptscriptstyle \pm 0.000}}$\\
\midrule

\multirow{3}{*}{Twitter}
& Real Data   & 0.607 & 0.902 & 0.950 & 0.404 & 0.133 & 1.000\\
\cmidrule(lr){2-8}
& w/o Profile & $0.287_{\scriptscriptstyle \pm 0.015}$ &  $0.339_{\scriptscriptstyle \pm 0.018}$ & $0.570_{\scriptscriptstyle \pm 0.020}$ & $0.146_{\scriptscriptstyle \pm 0.003}$ & $0.003_{\scriptscriptstyle \pm 0.000}$ & $0.566_{\scriptscriptstyle \pm 0.001}$\\
& w/ Profile  & $\mathbf{0.333_{\scriptscriptstyle \pm 0.014}}$ & $\mathbf{0.507_{\scriptscriptstyle \pm 0.018}}$ & $\mathbf{0.732_{\scriptscriptstyle \pm 0.024}}$ & $\mathbf{0.217_{\scriptscriptstyle \pm 0.010}}$ & $\mathbf{0.062_{\scriptscriptstyle \pm 0.011}}$ & $\mathbf{0.640_{\scriptscriptstyle \pm 0.006}}$\\
\bottomrule
\end{tabular}%
}
\end{table}

\subsubsection{Text Generation} 
Table~\ref{tab:linguistic_diversity} reports the linguistic diversity and semantic fidelity of text generated by Qwen2.5-14B, while similar results are observed for Llama-3.1-8B and omitted due to space limitations. Compared with the w/o Profile baseline, incorporating user profiles consistently improves lexical, semantic, and syntactic diversity, as well as semantic similarity, bringing generated comments closer to real user data. This suggests that personalized user modeling plays an important role in generating realistic and diverse content during the diffusion process.

\begin{figure}[t]
  \centering
  \makebox[\linewidth][c]{%
  \begin{minipage}{0.43\linewidth}
    \centering
    \includegraphics[height=2.4cm,keepaspectratio]{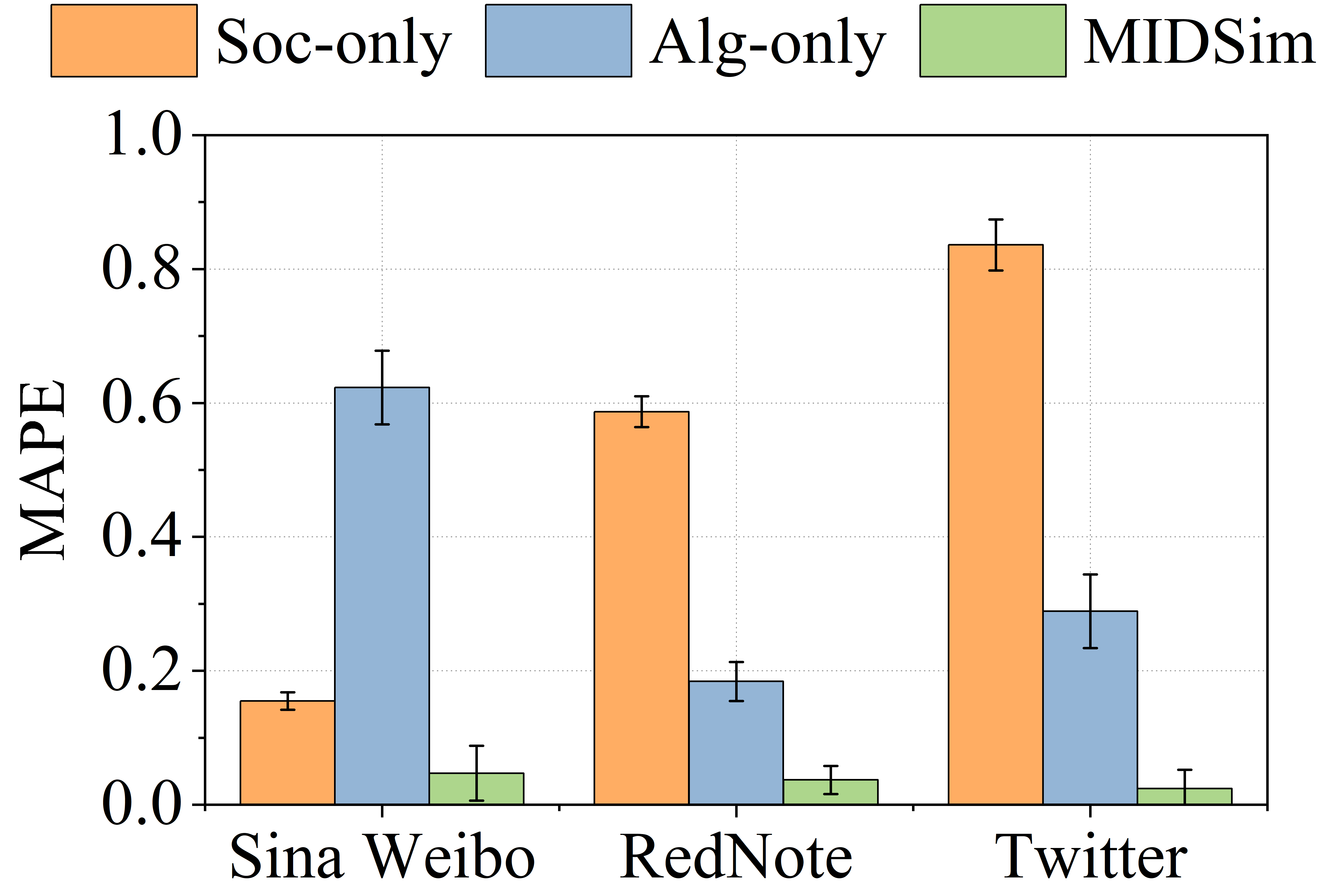}
    \vspace{-0.5em}
    \centerline{\small (a) Exposure channels}
  \end{minipage}
  \hspace{0.04\linewidth}
  \begin{minipage}{0.43\linewidth}
    \centering
    \includegraphics[height=2.4cm,keepaspectratio]{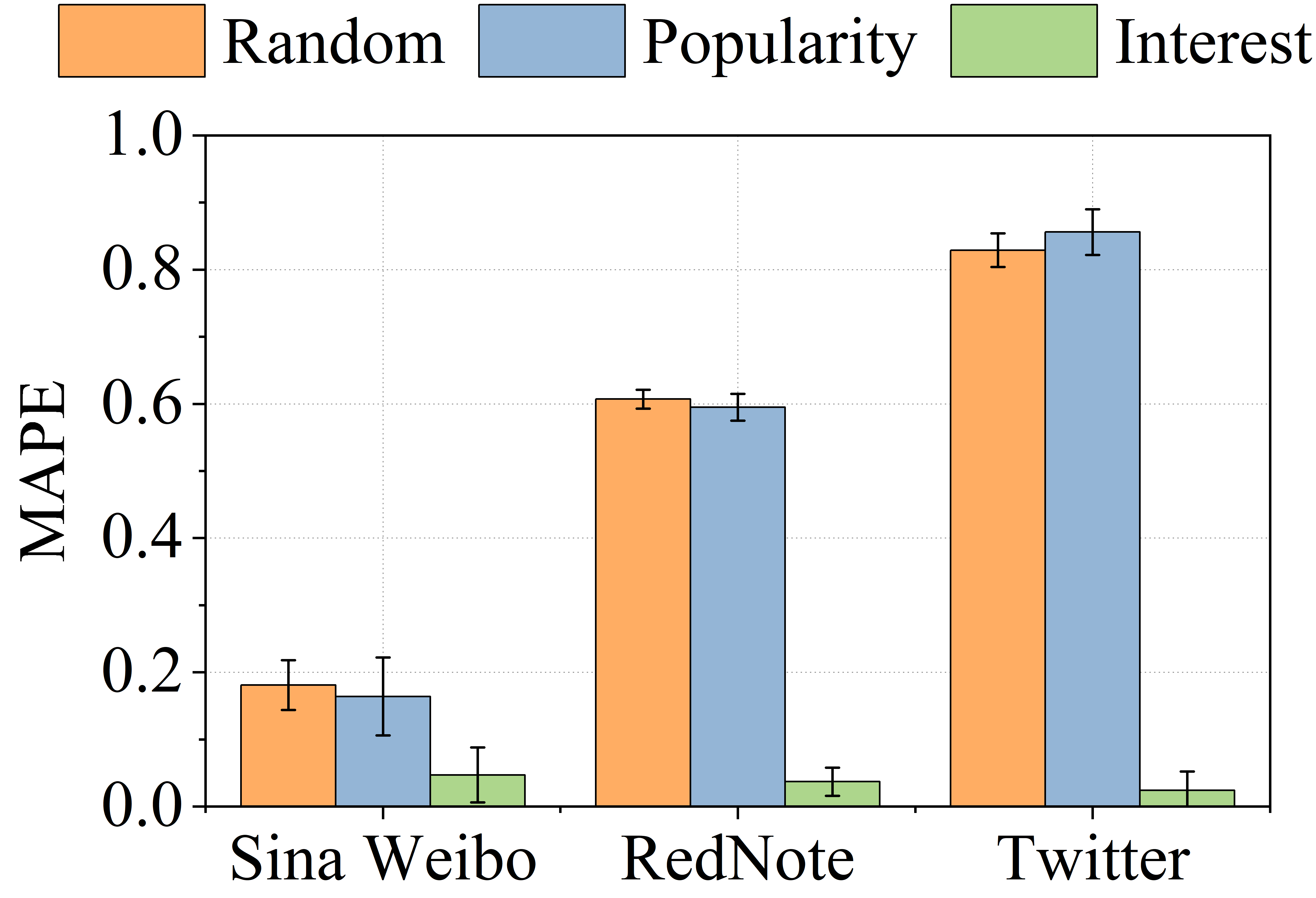}
    \vspace{-0.5em}
    \centerline{\small (b) Recommendation strategies}
  \end{minipage}%
  }
  \caption{Ablation study with Qwen2.5-14B.}
  \label{fig:channel_strategy_mape}
\end{figure}

\subsection{Ablation Study}
\subsubsection{Effect of Exposure Channels}
To analyze the contribution of different exposure channels, we compare Soc-only (follow feeds and notifications), Alg-only (recommendation and search feeds), and MIDSim (the full framework with both channels). As shown in Figure~\ref{fig:channel_strategy_mape}(a), MIDSim consistently achieves the best performance across all platforms, highlighting the complementary roles of social and algorithmic diffusion channels.
Among the single-channel variants, Soc-only performs better on Sina Weibo, indicating the important role of social exposure.
In contrast, Alg-only outperforms Soc-only on RedNote and Twitter, suggesting a stronger influence of algorithmic exposure on these platforms. 
 
\subsubsection{Effect of Recommendation Strategies} 
We further examine how different recommendation strategies affect diffusion simulation by replacing the default interest-based recommendation with popularity-based and random recommendation. As shown in Figure~\ref{fig:channel_strategy_mape}(b), interest-based recommendation consistently achieves the best performance. This finding highlights the importance of interest alignment in information diffusion and reflecting the interest-driven nature of modern recommendation systems.

\begin{figure}[h]
  \centering
  \includegraphics[width=0.9\linewidth]{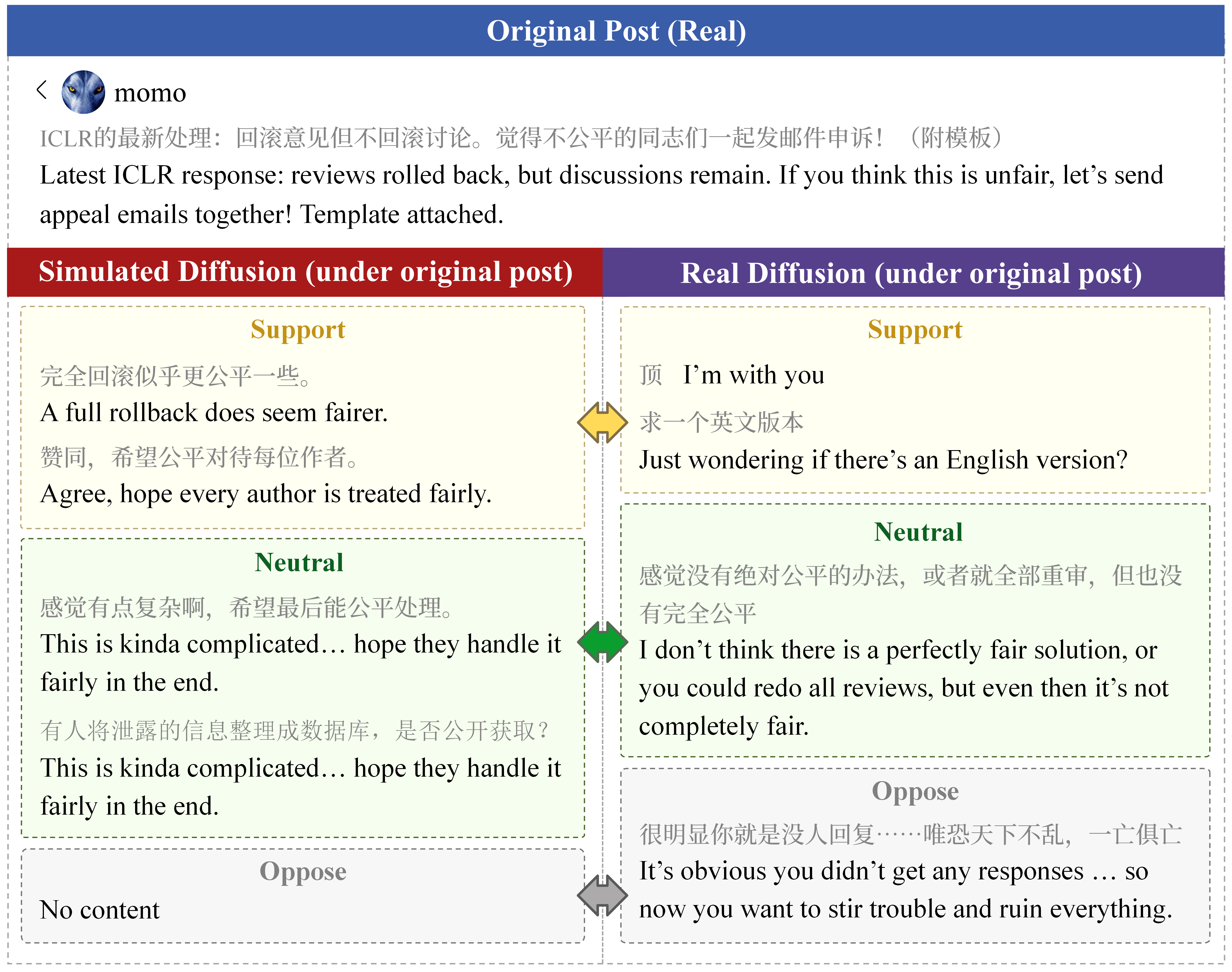}
  \caption{Case study of simulation under a RedNote post.}
  \label{fig:case_study}
\end{figure}

\subsection{Case Study} 
Figure~\ref{fig:case_study} presents a case study of a RedNote post, comparing simulated and real diffusion examples. Since the original content is in Chinese, English translations are provided for illustration. 
The simulated comments largely align with real diffusion in stance, primarily expressing supportive or neutral attitudes, while missing some opposing views observed in the real data. This observation suggests a tendency of current LLM-based agents to favor agreeable responses, highlighting the importance of better modeling dissenting and opposing viewpoints in future research.

\section{Conclusions}
In this paper, we propose MIDSim, an LLM-powered multi-agent framework for simulating multi-channel information diffusion that jointly models social and algorithmic exposure. To support realistic evaluation, we construct three real-world diffusion datasets spanning Sina Weibo, RedNote, and Twitter. Extensive experiments on real diffusion events demonstrate the effectiveness of MIDSim, highlighting the potential of LLM-driven agents for modeling complex information propagation. Future work includes broader cross-platform evaluation and more realistic modeling of user behaviors, recommendation mechanisms, and interaction diversity.

\bibliographystyle{ACM-Reference-Format}
\bibliography{sample-base}

\end{document}